\documentclass[12pt]{article}
\usepackage[latin1]{inputenc}      
\usepackage{graphicx}
\usepackage{epsfig}
\usepackage{amssymb}
\usepackage{pstricks}

\begin{document}

\begin{center}
\Large Monte Carlo simulation with fixed steplenght for diffusion processes in nonhomogeneous media
\end{center}

\begin{center}
\textit{V. Ruiz Barlett, M. Hoyuelos and H. O. Mártin}\\
Instituto de Investigaciones Físicas de Mar del Plata (CONICET-UNMdP)\\ and Departamento de Física, Facultad de Ciencias Exactas y Naturales,\\ Universidad Nacional de Mar del Plata,\\ Funes 3350, 7600 Mar del Plata, Argentina
\end{center}

\begin{abstract}
Monte Carlo simulation is one of the most important tools in the study of diffusion processes.  For constant diffusion coefficients, an appropriate Gaussian distribution of particle's steplengths can generate exact results, when compared with integration of the diffusion equation.  It is important to notice that the same method is completely erroneous when applied to non-homogeneous diffusion coefficients.  A simple alternative, jumping at fixed steplegths with appropriate transition probabilities, produces correct results.  Here, a model for diffusion of calcium ions in the neuromuscular junction of the crayfish is used as a test to compare Monte Carlo simulation with fixed and Gaussian steplegth.
\end{abstract}



\section{Introduction}

Diffusion processes with spatial dependent diffusion coefficient are not uncommon in physical, chemical or biological systems.  The literature on the subject is vast; we mention just a few examples.
 There is great interest in diffusion processes in a tube or channel of varying cross section \cite{Zwanzig,RegueraRubi,KalinayPercus,Berezhkovskii,RegueraRubi2,Burada,Berez,Martens}.  For a channel with cross section depending periodically on the longitudinal direction, it has been shown that the dynamics can be described with a one dimensional equation: the Fick-Jacobs approximation \cite{jacobs,Zwanzig}, and that this approximation can be improved with a spatial dependent diffusion coefficient \cite{RegueraRubi}. In \cite{AndersonLocalization} the authors demonstrated that Anderson Localization can be viewed as position-dependent diffusion. In \cite{DarkMatter}, the antiproton flux from dark matter annihilation-decay is studied using a position-dependent diffusion coefficient. In \cite{MusculoCorazon} a model of heart muscle is developed using a reaction-diffusion system with a spatially varying diffusion coefficient in order to investigate arrhythmia. In \cite{Lancon}, experimental results on brownian particles diffusing in a confined geometry are reported.  The particles are trapped between two nearly parallel walls.  They measured a spatial dependent diffusion coefficient and a drift in the direction of the diffusion coefficient gradient in the absence of an external force or concentration gradient. In \cite{Vaccaro} the gating in ion channels is described using a position-dependent stochastic diffusion model.

In \cite{Matveev}, a one-dimensional (1D) model for the description of diffusion of Ca$^{2+}$ in the neuromuscular junction of the crayfish was proposed, in which a fivefold increase of the diffusion coefficient takes place over a distance of around 100 nm.

Monte Carlo simulation is one of the most important tools in the study of diffusion processes.  It is known that, for a constant diffusion coefficient, the Monte Carlo (MC) method with a steplength taken from an appropriate Gaussian distribution gives exact results; this means that, on average, the result is the same than the one obtained from integration of the diffusion equation.  The situation is different for non-homogeneous systems.

The authors of \cite{Farnell} used the model of Ref.\ \cite{Matveev} to show that the MC method with Gaussian steplength and spatial dependent diffusion coefficient (i.e., spatial dependent width of the Gaussian distribution) leads to a systematic error, when MC averages are compared with the deterministic solution of the diffusion equation.  To correct this error, they proposed a modification of the Gaussian distribution whose main ingredient is to transform it into an asymmetric distribution in order to take into account the drift generated by the diffusion coefficient gradient.

In Ref.\ \cite{yo}, we proposed a different solution. We considered a discrete probability distribution for particle jumps at fixed distances to the left and right in a 1D lattice.  The jump length is constant over the whole system, and the jump probabilities are evaluated in terms of the diffusion coefficient and its gradient through simple relations.  The method applies to any smooth enough diffusion coefficient.  In Ref.\ \cite{yo}, we illustrated the method with a constant gradient diffusion coefficient.  Here, we apply it to the 1D model for diffusion of calcium ions in the neuromuscular junction of the crayfish \cite{Matveev} and compare the results with the ones obtained from MC simulation with Gaussian steplength and from numerical integration of the diffusion equation.  As already reported in \cite{Farnell}, the Gaussian steplength introduces an error that even increases with time.  The fixed steplength method proposed here gives a good agreement when compared with the deterministic results.

The model is used here as a test for fixed and Gaussian steplength in MC simulations.  Nevertheless, it also has biological interest. Diffusion of calcium ions plays an important role in chemical synapses.  Chemical synapses are specialized junctions through which neurons signal to each other and to non-neuronal cells such as those in muscles or glands.  The model includes a pulsed input of calcium ions generated by an action potential train (the input point is located in the voltage-gated calcium channel in the axon terminal).

The paper is organized as follows. In Sect.\ \ref{descrip} we present the general continuous and discrete description of the system.  In Sect.\ \ref{model} we describe the particular model and the results.  In Sect.\ \ref{concl} we state our conclusions.

\section{Continuous and discrete descriptions}
\label{descrip}

\subsection{Continuous description}

The connection between the diffusion or Fokker-Planck equation and the stochastic motion of Brownian particles is simple when the diffusion coefficient is homogeneous.  The problems grow when spatial dependence of the diffusion coefficient is taken into account.  Let us consider a 1D system.  The simplest Langevin equation for a Brownian particle, in position $x$, time $t$, and without a deterministic force (pure diffusion) is
\begin{equation}
\frac{dx}{dt} = \sqrt{2 D(x)} \zeta(t),
\end{equation}
where $D(x)$ is the diffusion coefficient and $\zeta(t)$ is the Gaussian white noise assumed to be $\delta$-correlated.  The corresponding Fokker-Planck equation for the concentration of particles $c(x,t)$ is \cite{sokolov}
\begin{equation}
\frac{\partial c}{\partial t} = \frac{\partial}{\partial x}\left[ \alpha \frac{\partial D}{\partial x} c + D \frac{\partial c}{\partial x} \right],
\label{fpgen}
\end{equation}
where $\alpha$ is the interpretation parameter.  The simplest interpretations (but not the only ones) are the ones of Ito ($\alpha=1$), Stratonovich ($\alpha=1/2$) and H\"anggi-Klimontovich ($\alpha=0$) \cite{sokolov}. It is clear from Eq.\ (\ref{fpgen}) that the interpretation parameter is relevant only when $D$ depends on $x$.

We consider the standard kinetic equation that is derived from the combination of continuity equation and Fick's law,
\begin{equation}
\frac{\partial c}{\partial t} = \frac{\partial}{\partial x}\left[D \frac{\partial c}{\partial x} \right],
\label{fpstandard}
\end{equation}
that corresponds to $\alpha=0$.  In this case, the equilibrium solution is homogeneous.  In a microscopic level this situation can be represented by diffusion in a potential where all potential wells are at the same level and the spatial dependence of $D$ is originated in variations of barrier's heights (barrier model).  This spatial dependence produces a drift velocity given by
\begin{equation}
v=\frac{\partial D}{\partial x}
\label{vdrift}
\end{equation}
(see Ref.\ \cite{yo}).

\subsection{Discrete description for MC simulation}

A usual choice to simulate diffusion processes is MC method with Gaussian steplength: in each MC step time is increased by $\Delta t$ and particles jump a distance given by $\sqrt{2 D(x) \Delta t} \xi$, where $\xi$ is a Gaussian stochastic variable with mean zero and dispersion 1.
As mentioned before, the MC simulation with Gaussian steplength produces exact results when the diffusion coefficient is homogeneous.  On the other hand, also in the homogeneous case, the fixed steplength choice has some important good features. It produces exact results in the evaluation of first and second order moments \cite{yo} (scaled errors in higher order moments behave as $t^{-1}$ for $t$ large).  And, more importantly, it keeps these good features when the diffusion coefficient is non homogeneous, in contrast with the Gaussian steplength.

\begin{figure}
\includegraphics{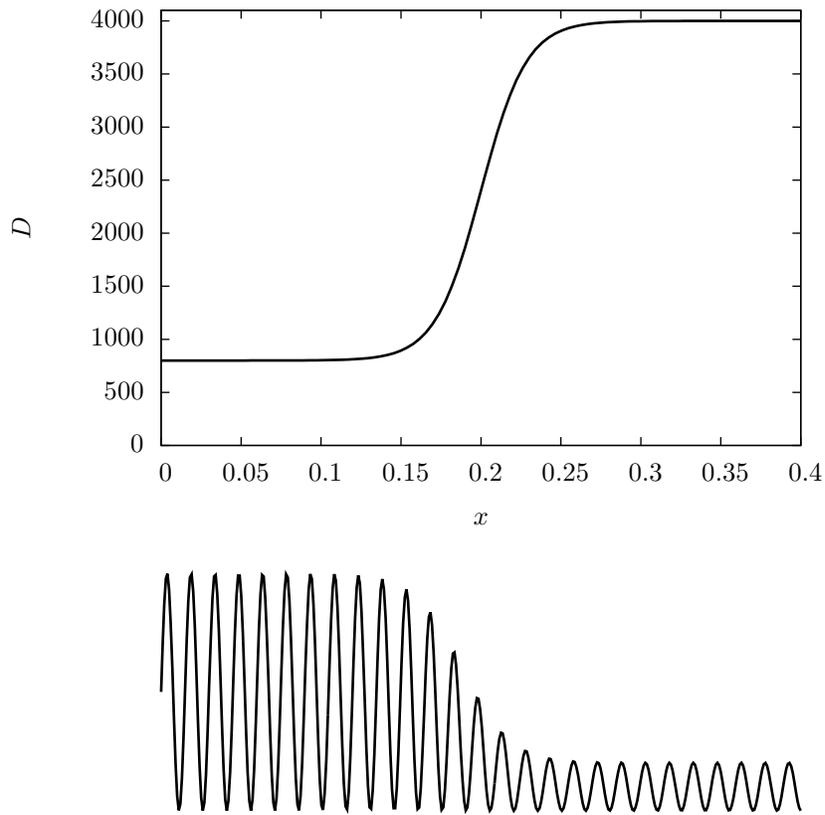}
\caption{Top: diffusion coefficient $D$ [nm$^2$/ms] of calcium ions against position $x$ [$\mu$m]. In the fixed steplength simulation the continuous space is replaced by a discrete lattice with energy barriers.  Bottom: the barriers are shown schematically (actual distance between sites is much smaller in the simulations).}\label{barreras}
\end{figure}

To implement the fixed steplength simulation, we consider a one dimensional discrete lattice with sites separated by potential barriers of different heights.  For example, Fig. \ref{barreras}  shows the spatial dependence of the diffusion coefficient of calcium ions in the neuromuscular junction of the crayfish \cite{Matveev}, and, below, we show schematically the corresponding barrier model for diffusion in a lattice.
Position of site number $i$ is $x_i = i\, \Delta x$, where $\Delta x$ is the (fixed) lattice spacing.  In each MC step, a particle in position $x_i$ can jump to the right or to the left with probabilities $p(x_i)$ and $q(x_i)$ respectively, and time is increased by a small time step $\Delta t$.  The transition probabilities per unit time are $P(x_i) = p(x_i)/\Delta t$ and $Q(x_i) = q(x_i)/\Delta t$.  Using the corresponding Master equation for these transition processes, we can arrive to a discretized version of the diffusion equation (\ref{fpstandard}), from which we can write the relations between transition probabilities and diffusion coefficient (see Eqs.\ (7) and (8) of Ref.\ \cite{yo}):
 \begin{equation}\label{px}
P(x_i) = \frac{{D(x_i)}}{{\Delta x^2 }} + \frac{1}{2\Delta x}\frac{\partial D}{\partial x}(x_i)
\end{equation}
\begin{equation}\label{qx}
Q(x_i) = \frac{{D(x_i)}}{{\Delta x^2 }} - \frac{1}{2\Delta x}\frac{\partial D}{\partial x}(x_i).
\end{equation}
The average drift velocity can be obtained from these equations:
\begin{equation}
v = \left[P(x_i)-Q(x_i)\right]\Delta x = \frac{\partial D}{\partial x}(x_i),
\label{vel}
\end{equation}
that is in agreement with Eq.\ (\ref{vdrift}).
The asymmetry between right and left jumps produces the drift current.  The main problem with the Gaussian distribution of steplengths is its symmetry, that precludes the appearance of the drift current.

\section{Model and results}
\label{model}

The model for the diffusion of clacium ions in the neuromuscular junction of the cryfish introduced in \cite{Matveev}, and used in \cite{Farnell}, was proposed in order to represent some experimental results for which it was necessary to assume a fivefold increase in the diffusion coefficient over a distance of the order of 100 nm.  The diffusion coefficient is
\begin{equation}\label{Dmatveev}
    D(x) = \hat D[1 - 0.8\, u(x)]
\end{equation}
\begin{equation}\label{ux}
u(x) = \frac{1}{2}\left\{ {\tanh \left[ {A\left( {b - x} \right)} \right] + 1} \right\},
\end{equation}
where $\hat D=4\times10^{-3}$ $\mu$m$^2$/ms, $A=35$ $\mu$m$^{-1}$ and $b=0.2$ $\mu$m (see Fig. \ref{barreras}).
The point where ions are injected in the system is $x=0$, this is where the voltage-gated calcium channel is located.  An action potential train of frequency 100 Hz opens the channel and, for each pulse, 8000 particles of calcium ions enter the system.  Particles are released during a span of 1.2 ms. The boundary condition at $x=0$ is reflecting. 

For the MC simulation with Gaussian or fixed steplength we used $\Delta t=0.00015$ ms.  During the 1.2 ms that lasts a pulse of the action potential, one particle enters the system in each MC step, so that 8000 particles enter for each pulse. Results were averaged over 50 samples.

For fixed steplength we used a lattice where position $x_i$ is given by $i \Delta x$ with $i=0,...,10^4$ and $\Delta x = 0.002$ $\mu$m (the lattice is large enough to avoid boundary effects).  The value of $\Delta x$ should be much smaller than any characteristic length of the system (in our case, around 0.1 $\mu$m) and, combined with $\Delta t$, they should satisfy $p(x_i)+q(x_i)<1$ $\forall i$.  For cases where jumping probabilities are very small in parts of the system, it is convenient to combine the fixed steplength with the exponential distribution of time steps given by kinetic MC to reduce computational time \cite{RuizB}.

We assume that the system is exactly described by the diffusion equation (\ref{fpstandard}).  The accuracy degree of the MC simulation results is determined comparing them with numerical integration of (\ref{fpstandard}).  Numerical integration was performed using the FTCS method \cite{Press}.
The initial condition is $c(x,0)=0$, and the discretization for numerical integration is also $\Delta x=0.002$ $\mu$m.

 \begin{figure}
  \includegraphics[scale=0.27]{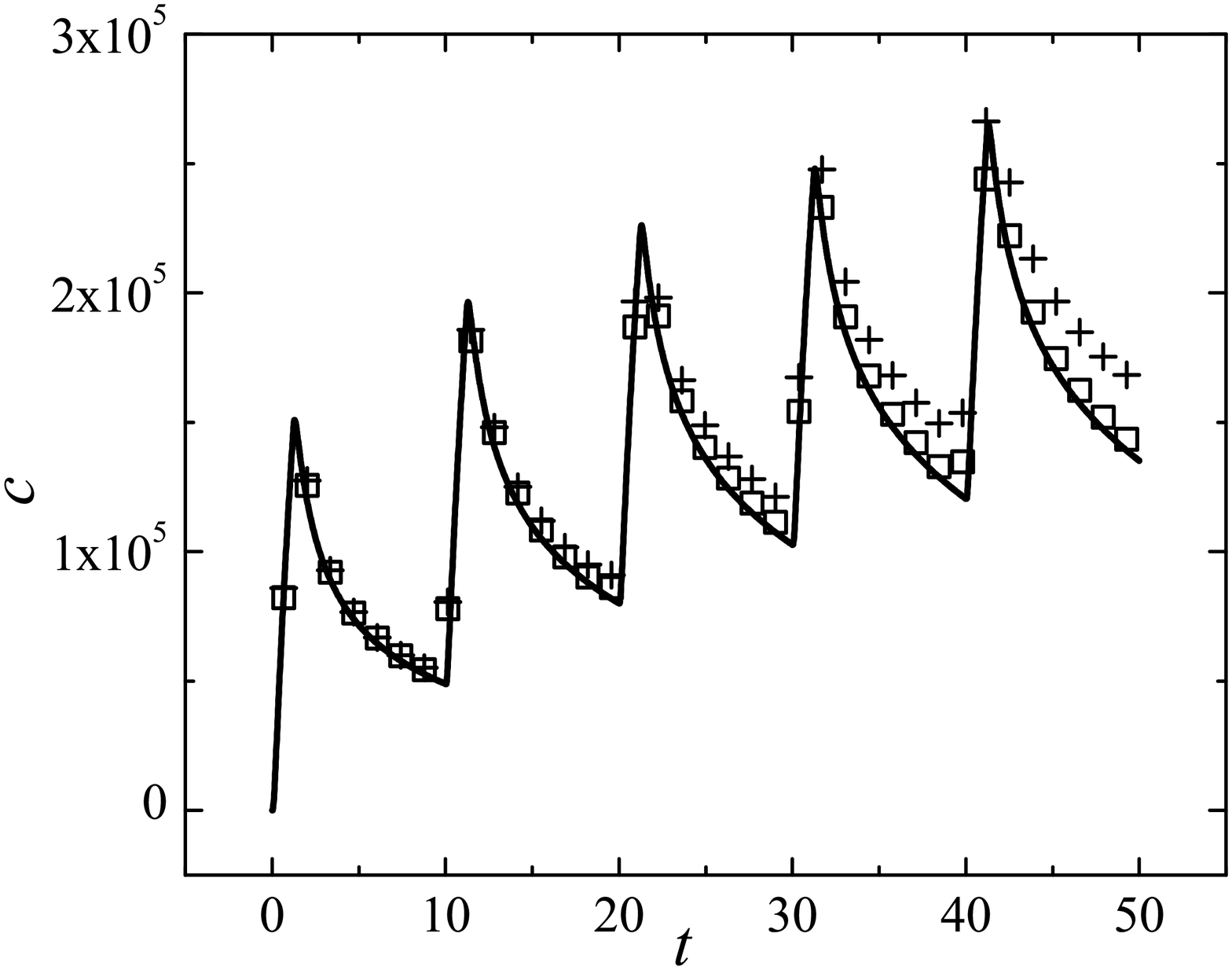}
  \includegraphics[scale=0.27]{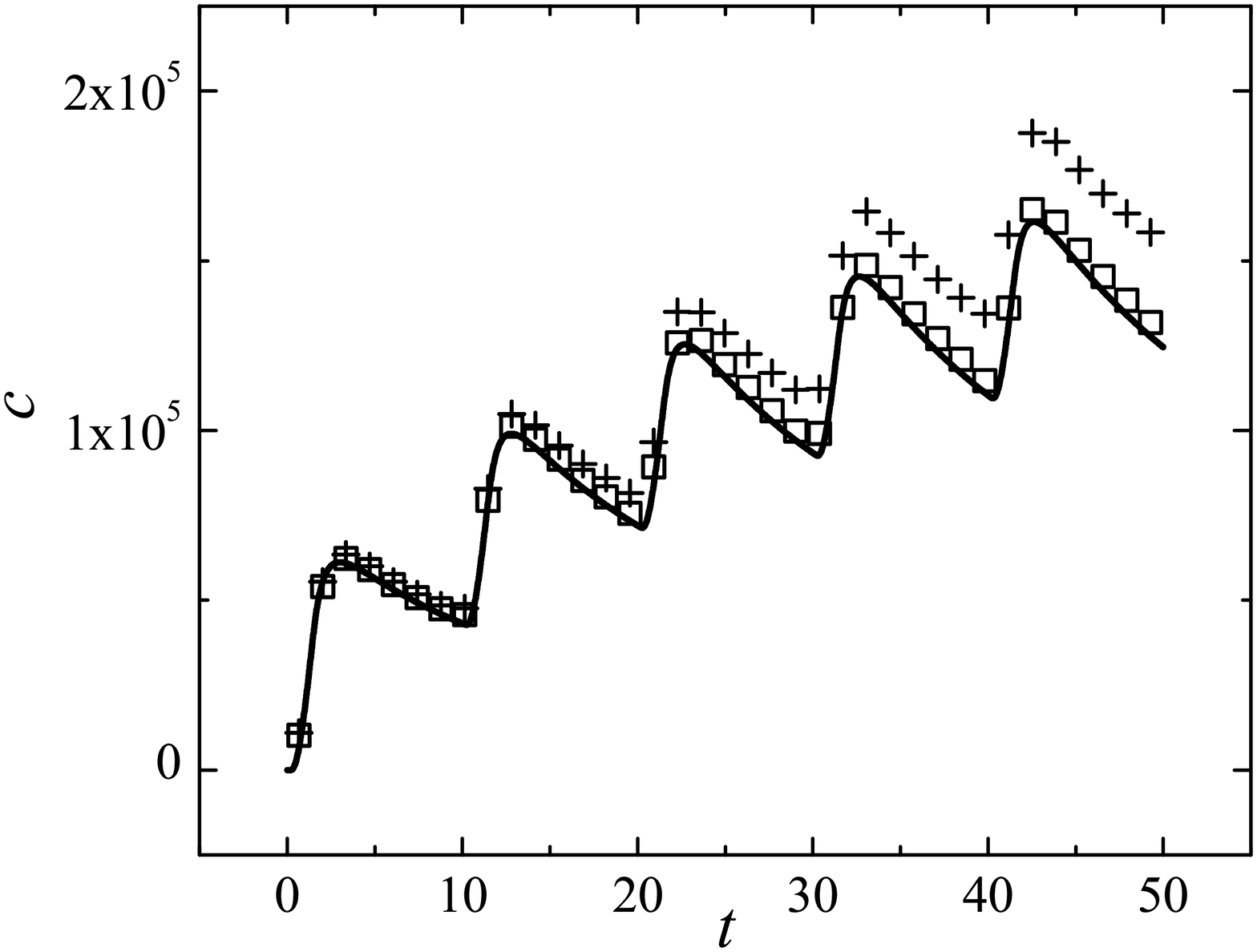}
 \vfill
  \includegraphics[scale=0.27]{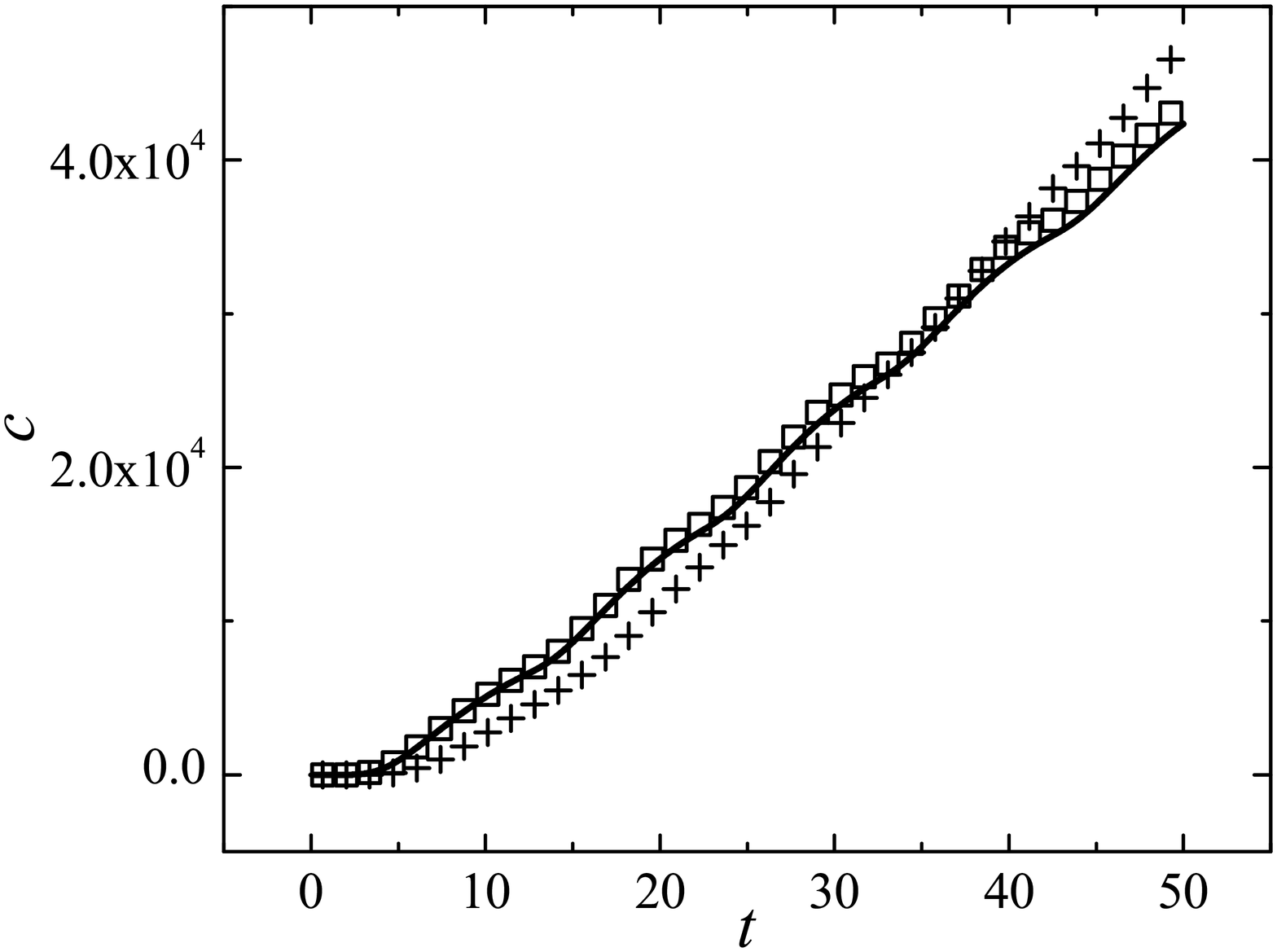}
  \includegraphics[scale=0.27]{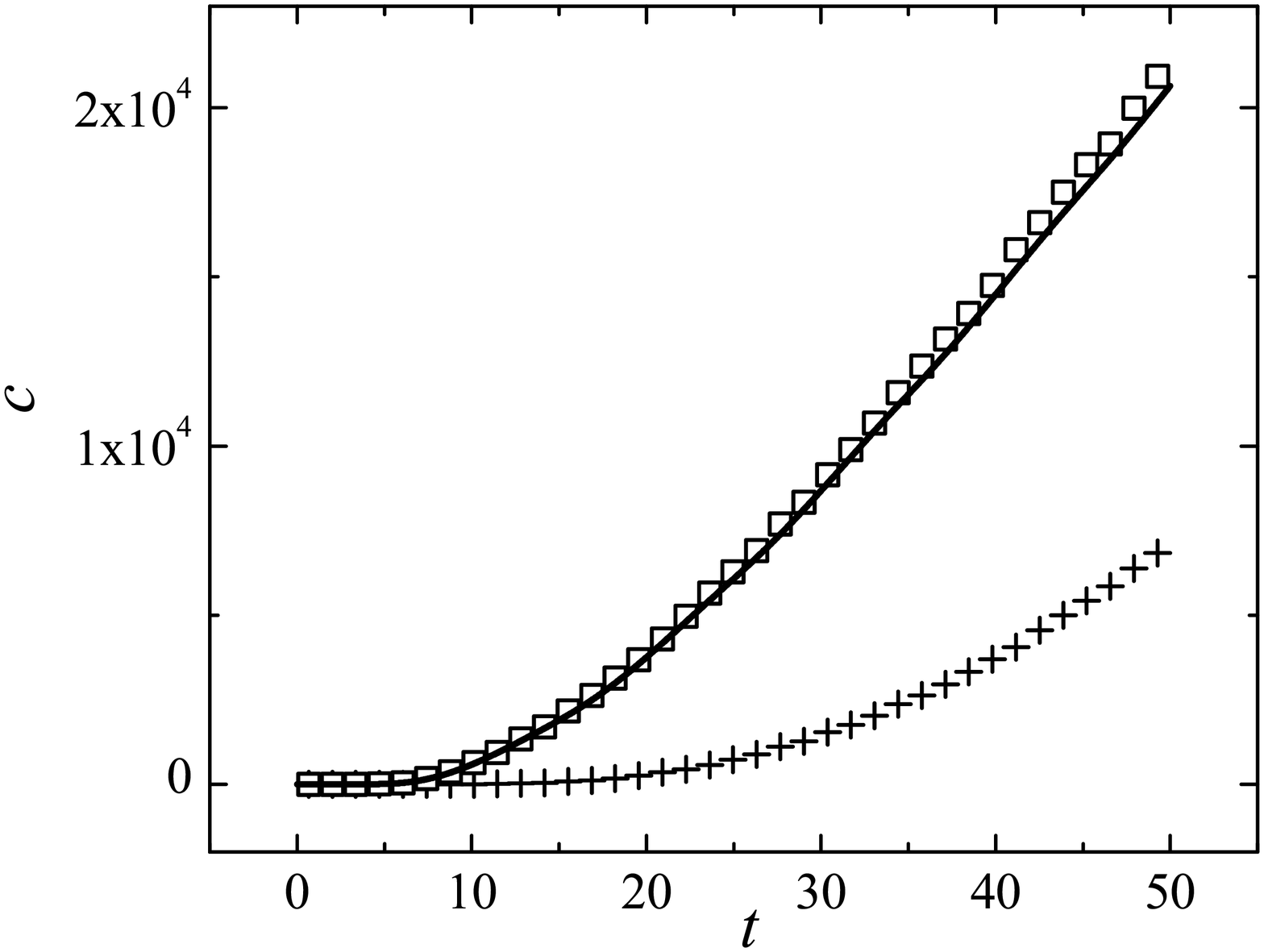}\\
  \caption{Concentration of particles $c$ (number of particles per $\mu$m) against time $t$ [ms] evaluated at different distances $x$ to the right of the input site. From left to right and from top to bottom: $x$=0.02, 0.06, 0.3 and 0.55 $\mu$m. Squares: fixed steplength; continuous curve (over squares): numerical integration of diffusion equation; plus signs: Gaussian steplength.}\label{cvst-distancias}
\end{figure}

 \begin{figure}
  \includegraphics[scale=0.27]{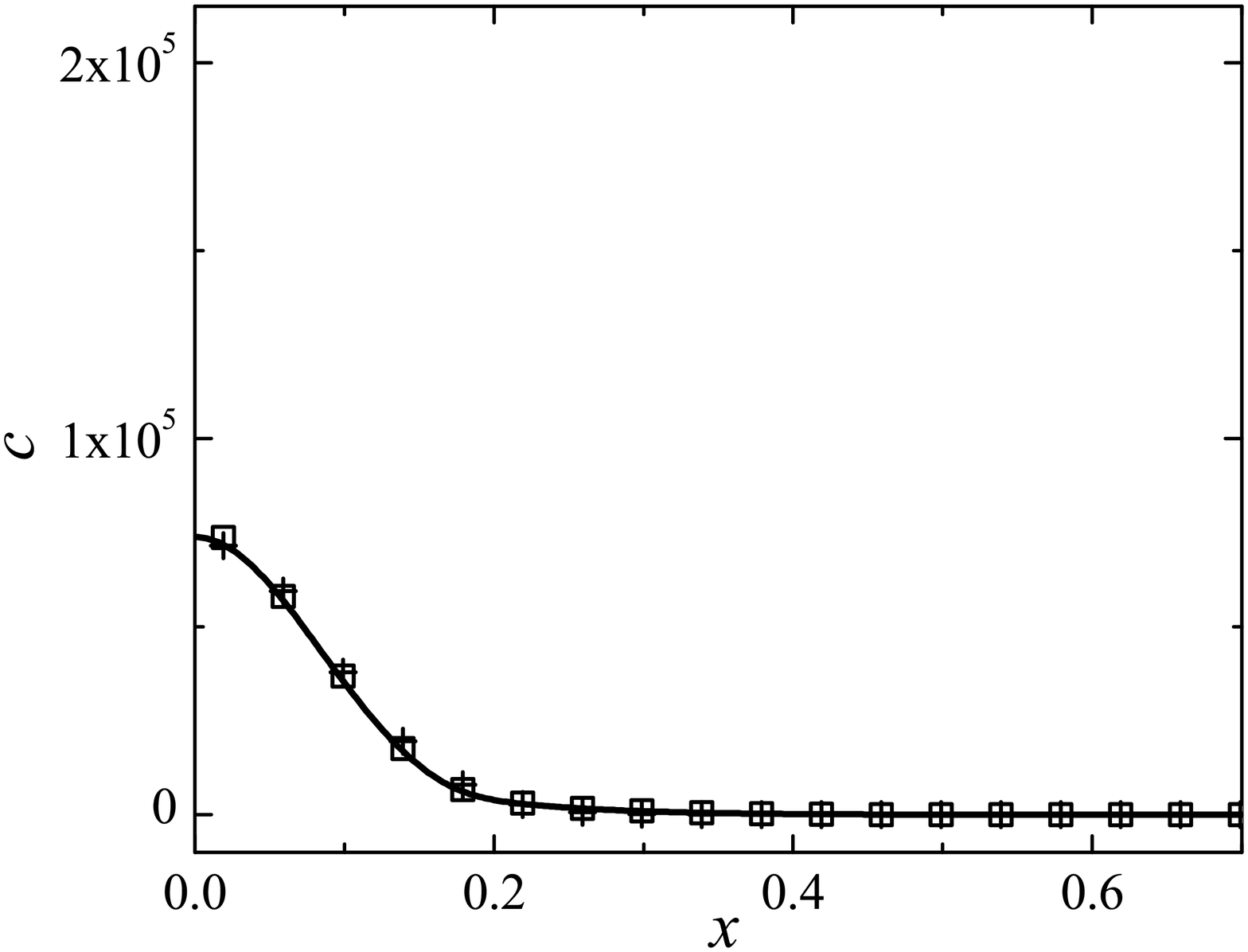}
  \includegraphics[scale=0.27]{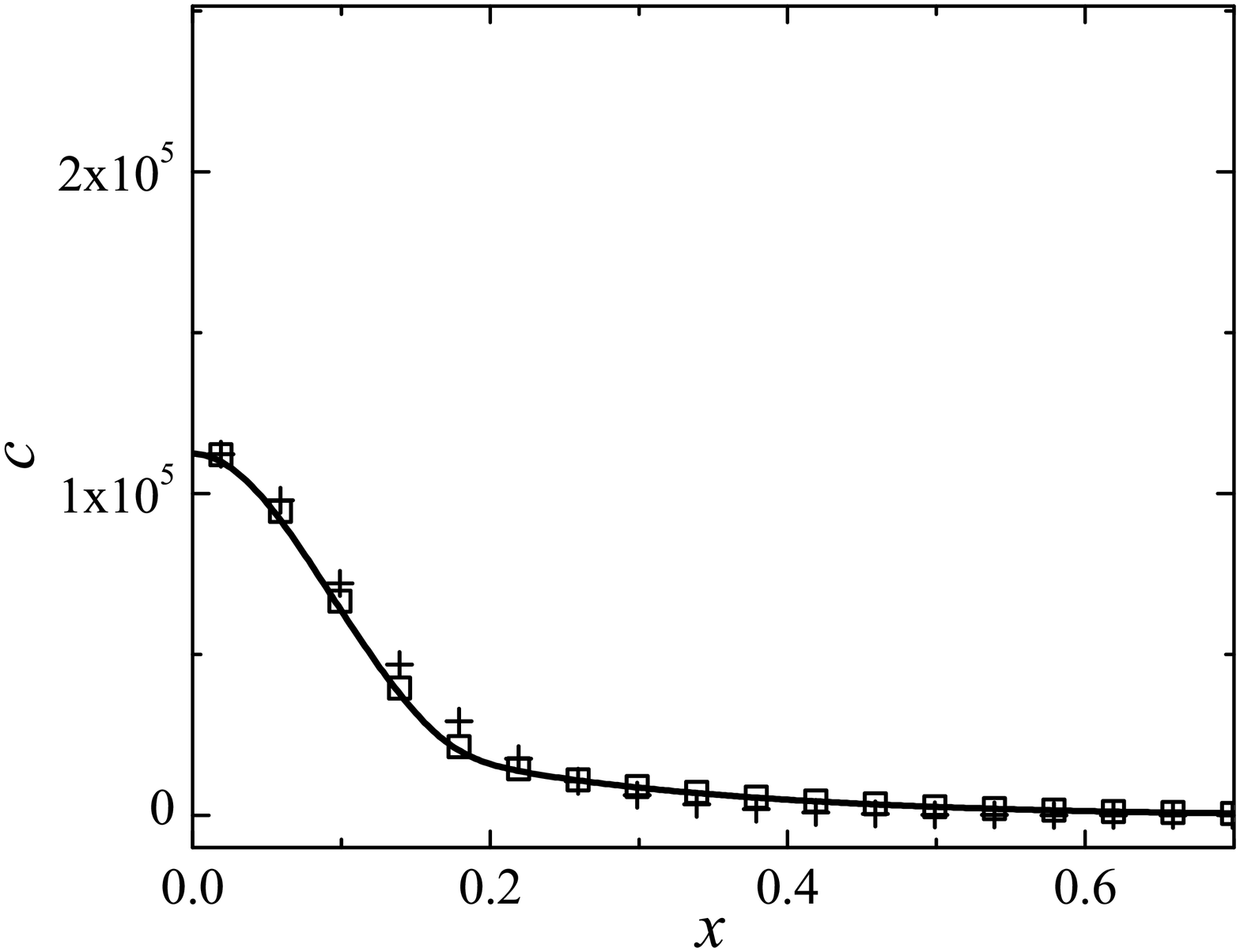}
 \vfill
  \includegraphics[scale=0.27]{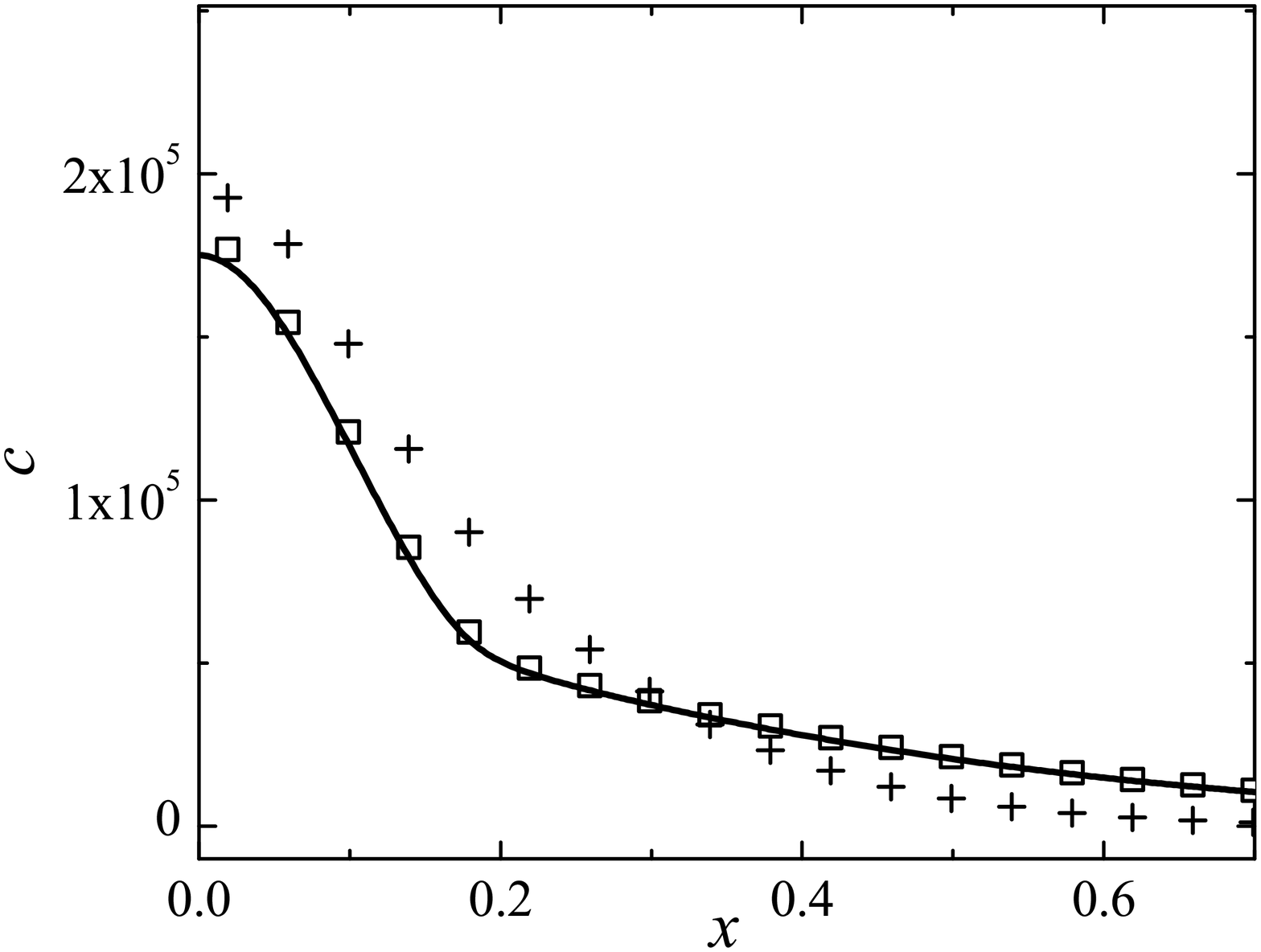}
  \includegraphics[scale=0.27]{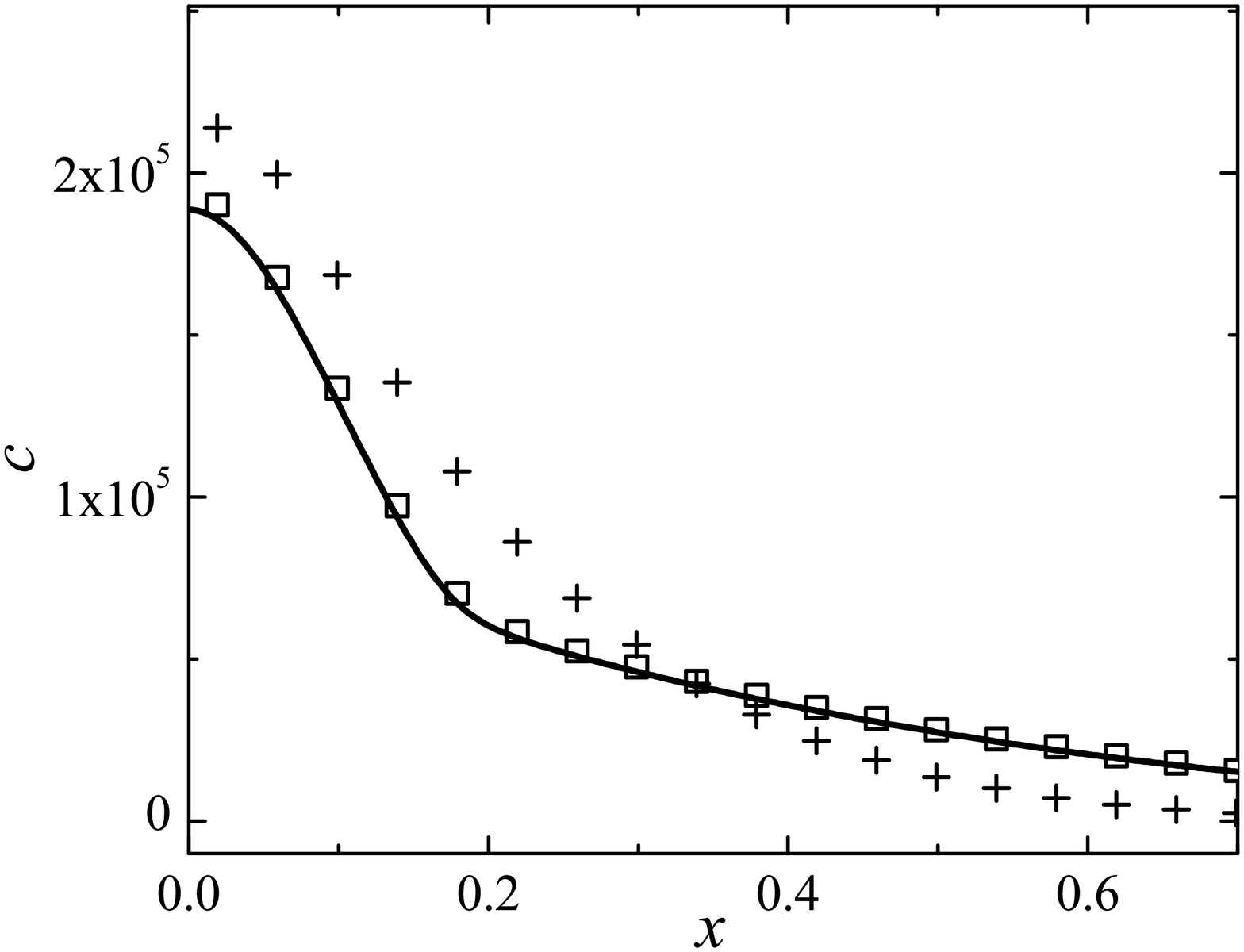}\\
  \caption{Concentration $c$ (number of particles per $\mu$m) against position $x$ [$\mu$m] for different times $t$ [ms]. From left to right and from top to bottom: $t=5$, 15, 45 and 55 ms. Squares: fixed steplength; continuous curve (over squares): numerical integration of diffusion equation; plus signs: Gaussian steplength.}
\label{cvsx-tiempos}
\end{figure}

Fig. \ref{cvst-distancias} shows the particle concentration as function of time evaluated at different distances $x$ to the right of the input site ($x=0.02$, 0.06, 0.3 and 0.55 $\mu$m). Numerical results with fixed steplength coincide with the results of integration of the differential equation for all times and distances. Gaussian steplength results do not coincide, and the error increases with time. The error is positive for short distances and negative for large distances. This fact can be understood analyzing the effect of the drift current.  The spatial dependence of $D$ (Fig. \ref{barreras}) induces a positive drift velocity, see Eq.\ (\ref{vdrift}).  For Gaussian steplength the drift is absent, and this produces an excess of particles close to the input site ($x$ small) and a lack of particles far from it to the right ($x$ large).

The difference between fixed and Gaussian steplength, and the absence of drift in the last one, is clearer in Fig.\ \ref{cvsx-tiempos}, where particle concentration against position for different times $t$ ($t=5$, 15, 45 and 55 ms) is plotted. The times were chosen in the middle of consecutive pulses.  As before, fixed steplength results coincide with the deterministic result of the diffusion equation, while Gaussian steplength results do not coincide.
The curve of fixed steplength results tends to favour larger values of $c$ for large $x$ because of the positive drift in the region around $x=0.2$ $\mu$m. For example, in the plot for $t=45$ ms or 55 ms, we can clearly see that the Gaussian steplength curve has an excess (lack) of particles for $x$ small (large), as mentioned before.

\begin{figure}
  \includegraphics[scale=0.5]{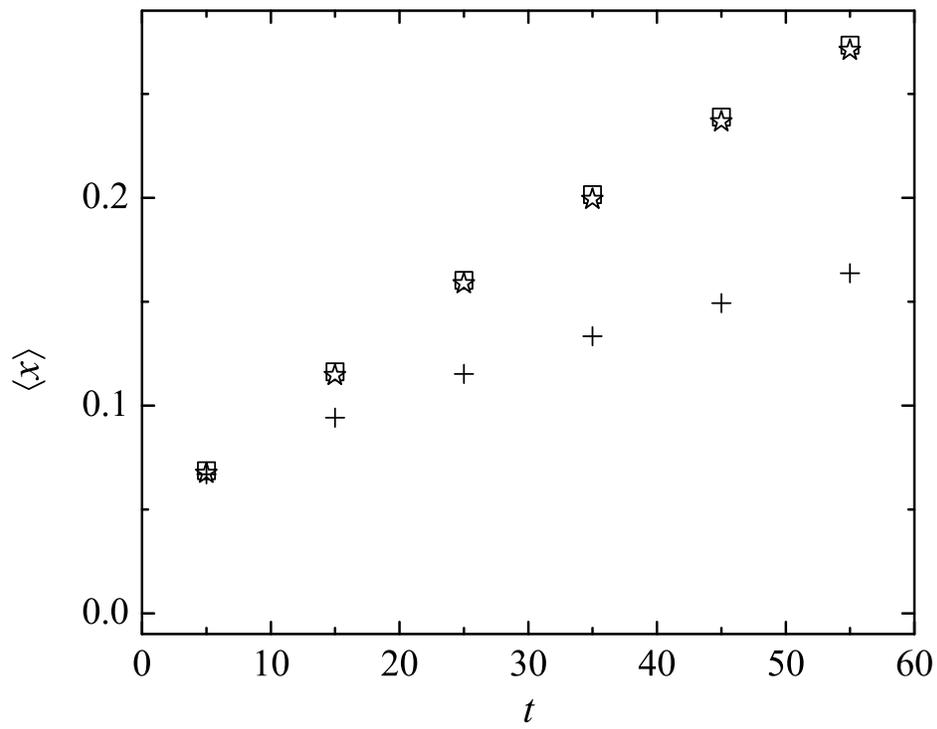}\\
  \caption{Average position $\langle x \rangle$ [$\mu$m] against time [ms].  Squares: fixed steplength; stars (over squares): numerical integration of diffusion equation; plus signs: Gaussian steplength.}\label{xm}
\end{figure}

The effect of the absence of drift in the Gaussian steplength results is also present in the evaluation of the average position $\langle x \rangle$ against time, as shown in Fig.\ \ref{xm}.  It can be seen that the results of $\langle x \rangle$ for fixed steplength and numerical integration coincide. But, for Gaussian steplength, $\langle x \rangle$ is smaller for all $t$. In both cases, $\langle x \rangle$ increses with time due to the continuous input of particles and the reflecting boundary condition at $x=0$.  The difference between Gaussian and fixed steplength is due to the drift velocity.  Particle input is actually pulsed, but pulses are not present in this plot because time was chosen in the middle of consecutive pulses.

\section{Conclusions}
\label{concl}

We have used a model for diffusion of calcium ions in the neuromuscular junction of the crayfish to compare MC simulation results with fixed and Gaussian steplength.  The system is assumed to be exactly described by the diffusion equation.  Comparison of the MC simulation results with numerical integration of the diffusion equation shows a time increasing error for the Gaussian steplength choice, while fixed steplength coincide with numerical integration.  Results of the particle density at different times and at different positions for Gaussian steplength show errors that can be understood in terms of the symmetric shape of the Gaussian distribution, that can not produce the drift generated by the diffusion gradient.  This feature makes the Gaussian steplength completely inadequate for the simulation of non-homogeneous systems.

In a previous work \cite{yo}, we have shown that the fixed steplength method exactly reproduces first and second order moments in homogeneous systems, and that the scaled error in higher order moments behaves as $t^{-1}$ for $t$ large.

The method is proposed as a simpler alternative to the methods of Refs. \cite{Farnell,Farnell2} to correct the errors introduced by the Gaussian steplength in non homogeneous systems.

\section*{Acknowledgments}

This work was partially supported by Consejo Nacional de Investigaciones Científicas y Técnicas (CONICET, Argentina, PIP 2010-2012 Nro. 0041) and Comisión de Investigaciones Científicas (CIC, Buenos Aires, Argentina).


\end{document}